\documentclass[11pt]{article}
\usepackage{moriond,epsfig}

\def\be{\begin{equation}}
\def\ee{\end{equation}}
\def\bea{\begin{eqnarray}}
\def\eea{\end{eqnarray}}

\newcommand{\lsim}   {\mathrel{\mathop{\kern 0pt \rlap
  {\raise.2ex\hbox{$<$}}}
  \lower.9ex\hbox{\kern-.190em $\sim$}}}
\newcommand{\gsim}   {\mathrel{\mathop{\kern 0pt \rlap
  {\raise.2ex\hbox{$>$}}}
  \lower.9ex\hbox{\kern-.190em $\sim$}}}

\newcommand{\pabar}{\not{\!\partial}}
\newcommand{\Od}{{\cal O}}

\newcommand{\Dbar}{\not{\!{\!D}}}

\begin{document}
\vspace*{4cm}
\title{PRODUCTION OF HIGH-ENERGY GRAVITINOS DURING PREHEATING}

\author{A.L MAROTO$^\dagger$ AND J.R. PELAEZ$^*$ }

\address{$^\dagger$CERN Theory Division, CH-1211 Geneva 23,
  Switzerland\\
$^*$Departamento de F\'{\i}sica Te\'orica, Universidad
  Complutense de Madrid, 28040 Madrid, Spain}

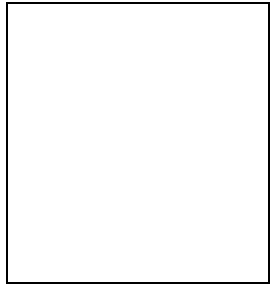
\maketitle\abstracts{
We present a new technique for the calculation of helicity $\pm 1/2$
gravitino production during preheating. It is based on the 
equivalence between goldstinos and  helicity $\pm 1/2$ gravitinos at
high energies. The problem is thus reduced to the standard 
(Majorana) fermion production after inflation. Comparison with
the results obtained in the unitary gauge is also presented.}

\section{Introduction}
From supersymmetry it is possible  to build models of
inflation in a relatively natural way. Since 
scalar fields appear as superpartners of 
fermionic matter fields,  there is no need to introduce
them {\it ad hoc} as in other models.
In addition, the potential energy of
those scalar fields typically contain flat directions which make
them natural inflaton candidates. Moreover, due to 
the nonrenormalization theorems, those flat directions are not
spoiled by radiative corrections. 

However, when promoting supersymmetry to a local symmetry
(supergravity), new problems may appear. First, it is known that
supergravity
corrections can modify the slow-roll parameter $\eta$, thus spoiling
the inflationary period (at least in minimal sugra models). 
Another problem that supergravity creates is related to the 
superpartner of the graviton field, the gravitino.  

Gravitinos are weakly interacting particles with very long lifetimes
(for a typical mass around $1$ TeV, gravitinos can live as
long as $10^5$ s). This implies that their  
decay products could destroy the nuclei created in the nucleosynthesis
period. This imposes very stringent constraints on the primordial
abundance of gravitinos. Thus, for instance, for $m_{3/2}\simeq 1$TeV,
the number density to entropy density ratio should satisfy 
$n/s\lsim 10^{-14}$.
Since gravitinos can be created after inflation
due to particle collisions in the thermal bath generated in the
reheating period, this constraint imposes the well-know bound on the
maximum reheating temperature $T_R \lsim 10^9$ GeV.

However, apart from particle collisions, gravitinos can be generated
directly from the inflation oscillations in the preheating period.
This production is much more efficient than the thermal one and
therefore much more dangerous for nucleosynthesis. Here we review
some of our recent results \cite{nosotros} on how to calculate
gravitino production during preheating, by means of
the high energy equivalence between goldstinos and helicity $\pm1/2$
gravitinos. For further details and references we refer the reader to that 
work \cite{nosotros}.

\section{Supergravity Lagrangian and gravitino helicities}

We will consider minimal supergravity coupled to a single chiral
superfield which contains an scalar field $\phi$ (inflaton) and
a Majorana spinor $\eta$ (inflatino, goldstino). We give only
the form of the fermionic Lagrangian up to quadratic terms in the
fields
\begin{eqnarray}
g^{-1/2}{\cal L}_F&=&-\frac{1}{2} \epsilon^{\mu\nu\rho\sigma}
\bar\psi_\mu \gamma_5 \gamma_\nu D_\rho
\psi_\sigma+\frac{i}{2}\bar \eta
\Dbar \eta +e^{G/2}\left( \frac{i}{2}\bar \psi_\mu \sigma_{\mu\nu}
\psi^\nu+\frac{1}{2}\left(-G_{,\phi\phi}-G_{,\phi}^2\right)
\bar \eta \eta \right. \nonumber \\
&+&\left. \frac{i}{\sqrt{2}}G_{,\phi}\bar \psi_\mu
\gamma^\mu\eta\right)+\frac{1}{\sqrt{2}}\bar \psi_\mu 
(\pabar \phi) \gamma^\mu
\eta,
\label{lagrangiano}
\end{eqnarray}
where the K\"ahler potential is given by
$G(\Phi,\Phi^\dagger)=\Phi^\dagger\Phi+\log \vert W \vert^2$.
Note that the last two terms contain mixing between gravitinos
$\psi_\mu$ and goldstinos, and therefore their equations of motion
are coupled. In addition, and in order to have a consistent model,
we will impose that the inflaton potential energy 
vanishes at the minimum, so that the cosmological constant is zero.
In addition, we will require 
the derivative of the K\"ahler potential $G_{,\phi_0}$ to be $\sqrt{3}$
 at the minimum $\phi=\phi_0$, i.e.
to be different from zero and therefore to have broken supersymmetry.
The above Lagrangian together with the corresponding  bosonic one
display a gauge invariance associated to local supersymmetric 
transformations. As usual, when the gauged supersymmetry is 
spontaneously broken, 
the gravitino field acquires a mass $m_{3/2}=e^{G_0/2}$, by means of
the 
super-Higgs
mechanism. Hence, since the gravitino now
is a  massive spin $3/2$ particle,  it also has states with $\pm 1/2$ 
helicity, 
in addition to the helicity $\pm 3/2$
states already present in the massless case.

As long as we are interested in the production of gravitinos during the
preheating era at the end of inflation, we will consider the equations
of motion, derived from the above Lagrangian, in a
Friedmann-Robertson-Walker background and when the inflaton is
a homogeneous field only depending on time. Then, it can
be seen that the equations of motion for $\pm 3/2$-helicity gravitinos
reduce to the Dirac-like equation \cite{gravi}
\begin{eqnarray}
(i\Dbar-e^{G/2})\psi_\mu^{\pm 3/2}=0
\label{g32}
\end{eqnarray}
Thus, the production of helicity $\pm 3/2$ gravitinos can be
treated as the production of Dirac fermions during preheating. 
However, the helicity $\pm 1/2$ equation is still coupled to the
goldstino and very complicated. Two approaches can be followed in this
case. In the first one\cite{Linde} the gauge is fixed by
imposing $\eta=0$, so that the
goldstino disappears from the Lagrangian, and we are
left with the equations of motion of a pure spin $3/2$ spinor. 
This gauge is called unitary since the 
unphysical degrees of freedom are not explicit. The 
second approach, which
is the one we are going to present here, is based on the 
so-called $R_\xi$ gauges, where it is possible to find a 
high energy relation
between the helicity $\pm 1/2$ gravitinos and goldstinos.
Thus we will be able to calculate the
helicity $\pm 1/2$ gravitino production from the much simpler
production of goldstinos.    

\section{Gauge fixing and the equivalence theorem}

Let us consider the following gauge condition:
\begin{eqnarray}
\gamma^\mu\psi_\mu
-\frac{1}{\sqrt{2}\xi\Dbar}e^{G/2}G_{,\phi}\eta
+\frac{i}{G_{,\phi}}e^{-G/2}\gamma^\mu(\pabar\phi)\psi_{\mu}=0.
\end{eqnarray}
where $\xi$ is an arbitrary parameter. 
We will assume in the following that the space-time is asymptotically
flat and that also asymptotically in $t\rightarrow \pm \infty$ the
inflaton field settles down at the minimum of the potential 
$\phi \rightarrow \phi_0$. With these conditions we see that in those
asymptotic $in, out$ regions, the above condition simplifies to
\begin{eqnarray}
a^{-1}_{in,out}\pabar\gamma^\mu\psi_\mu
=\sqrt{\frac{3}{2}}\frac{m_{3/2}}{\xi}\eta,
\end{eqnarray}
where $a_{in, out}$ are the asymptotic values of  the scale factor. 
If we use the equations of motion for gravitinos and goldstinos,
this gauge condition can be rewritten as
\begin{eqnarray}
\partial^\mu\psi_\mu=\sqrt{\frac{3}{2}}\frac{m}{\xi}
\left(1-\xi\frac{m_{\pm}}
{m}\right)
\eta,
\label{gg}
\end{eqnarray}
where we have redefined $m=a_{in,out}m_{3/2}$ and 
$m_{\pm}=m(1\pm \sqrt{1-3/(2\xi)})$. Note that
there are two different relations for goldstinos 
with masses $m_-$  and $m_+$, but 
the important point is that the {\em derivative} of the
gravitino field is proportional to the goldstino.

Let us then introduce the equivalence theorem. In the asymptotic
regions that we mentioned before it is expected that the general solution
of the equations of motion for gravitinos and goldstinos can be
written as linear superpositions of plane waves. In fact, at high
energies,
since the effects of the difference in masses between gravitinos and
goldstinos is negligible, the solutions are
\begin{eqnarray}
\psi_\mu^{p}(x)=\frac{1}{a^{3/2}\sqrt{2\omega}}
e^{i px}\tilde\psi_\mu(\vec p)
+\Od\left(\frac{m}{\omega}\right),\;\;\;
\eta^{p}(x)=\frac{1}{a^{3/2}\sqrt{2\omega}}e^{i px} 
\tilde \eta(\vec p)
+\Od\left(\frac{m}{\omega}\right),
\label{goldi}
\end{eqnarray}
where we assume $p_\mu p^\mu=m^2$ and $\omega \gg m$.
Again at high energies, the helicity $\pm 1/2$ projector for gravitinos is
\begin{eqnarray}
P_\mu^{\pm 1/2}=\sqrt{\frac{2}{3}}P_{\pm}\frac{p_\mu}{m}
+\Od\left(\frac{m}{\omega}\right),
\end{eqnarray}
so that, in momentum space, the helicity $\pm 1/2$ component
is given by
\begin{eqnarray}
\tilde\psi_{\pm 1/2}(\vec p) \equiv P^\mu_{\pm
  1/2}\tilde\psi_\mu(\vec p)=\sqrt{\frac{2}{3}}
P_{\pm}\frac{p^\mu}{m}\tilde\psi_\mu(\vec p)
+\Od\left(\frac{m}{\omega}\right).
\label{preTE}
\end{eqnarray}
We see that up to a correction negligible at
high energies, the helicity $\pm 1/2$ gravitino is proportional 
to $\partial^\mu\psi_\mu$, but from (\ref{gg}) this in turn is
proportional
to the goldstino field. Therefore we can write 
 \begin{eqnarray}
\tilde\psi_{\pm 1/2}(\vec p)= \sum_{+,-}
\left[-i\frac{1}{\xi}\left(1-\xi\frac{m_{+,-}}
{m}\right)P_{\pm 1/2}
+\Od\left(\frac{m}{\omega}\right)\right]\tilde\eta^{+,-}(\vec p).
\end{eqnarray}
This equation still relates the $\pm 1/2$ helicity gravitino
with the two different goldstino solutions, either with $m_-$ or $m_+$.
However, by choosing $\xi=3/2$ \cite{Casalbuoni} we obtain $m_-=m_+$
and there is a unique high energy
relation between $\pm1/2$ helicity gravitino and goldstinos.
Another, even simpler, possibility \cite{nosotros} is to choose 
the Landau gauge
$\xi\rightarrow\infty$, where there is only one $m_+$ solution. 
In this gauge, we can then write the equivalence theorem as
 \begin{equation}
\tilde\psi_{\pm 1/2}(\vec p)=\left[2\,i \,P_{\pm}+
\Od\left(\frac{m}{\omega}\right)\right]\eta(\vec p).
\label{TE}
\end{equation}

\section{Particle production}
The previous expression valid in the asymptotic regions is sufficient
to relate the production of helicity $\pm 1/2$ gravitinos to 
the production of goldstinos. In fact, let us consider a pure
positive(negative) frequency mode solution for goldstinos in the
$in$ region
\begin{eqnarray}
\eta^{p}_l(x)\rightarrow \frac{1}{a^{3/2}_{in}\sqrt{2\omega_{in}}}
e^{i\omega_{in} t -i\vec p \vec x}u(\vec p, l).
\end{eqnarray}
Because of the presence of the oscillating inflaton field and the
space-time
curvature, this solution will no longer behave as pure
positive(negative)
frequency mode in the $out$ region, but it will be a linear 
superposition of positive and negative frequency modes
\begin{eqnarray}
\eta^{p}_l(x)\rightarrow \frac{1}{a_{out}^{3/2}\sqrt{2\omega_{out}^+}}
\left(\alpha^G_{p,l} 
e^{i\omega_{out}^+ t -i\vec p \vec x}
u(\vec p, l) +\beta^G_{-p,l} e^{-i\omega_{out}^+ t -i\vec p \vec x}
u^C(-\vec p, l)\right),
\end{eqnarray}
where $\alpha^G_{p,l}$ and $\beta^G_{-p,l}$ are known as Bogolyubov 
coefficients. Since the Fourier modes of goldstinos are related to 
the Fourier modes of gravitinos in the asymptotic regions, using
(\ref{TE})
we can find a relation between the particle numbers of helicity 
$\pm 1/2$ gravitinos and goldstinos:
\begin{eqnarray}
N^L_{p,l}=\left[1+\Od\left(\frac{m}{p}\right)\right]\
 \vert \beta^G_{p,l} \vert ^2.
\label{number}
\end{eqnarray}
Thus in order to obtain the helicity $\pm 1/2$ 
gravitino production $N^L_{p,l}$,  
we only need to know the goldstino coefficients $\beta^G_{p,l}$. With
that purpose we have to solve the equation of motion for the
goldstinos. However  in the Landau gauge 
the equation of motion of the goldstinos reduces again to a Dirac-like
equation, in particular
\begin{eqnarray}
i\Dbar \eta - e^{G/2}\left(G_{,\phi\phi}+G_{,\phi}^2\right)\eta=0.
\label{goldstin}
\end{eqnarray}
This is an additional reason to use the Equivalence Theorem in 
the Landau gauge.
The problem of helicity $\pm 1/2$
production
is thus reduced as in the helicity $\pm 3/2$ case to a Dirac fermions
calculation.
In order to check these results, we can see that in the limit 
$\vert \phi \vert\ll  M_P$, the 
equation for the goldstinos in the Landau gauge (\ref{goldstin}) can
be approximated by
\begin{eqnarray}
i\Dbar \eta - \left(\partial_\phi \partial_\phi W\right)\eta=0,
\label{susy}
\end{eqnarray}
which is the equation obtained in the unitary gauge\cite{Linde} for the helicity
$\pm 1/2$ gravitinos (in the global supersymmetric
limit). Therefore,
the number of goldstinos, calculated in the Landau gauge, 
is the same as that of
helicity $\pm 1/2$ gravitinos, calculated in the unitary gauge. 
 
{\bf Acknowledgements:} This work has been partially supported by 
CICYT-AEN97-1693, PB-98-0782 and Ministerio de Educaci\'on y Ciencia (Spain).

\end{document}